\documentclass[pdftex,twocolumn,epjc3]{svjour3}          

\RequirePackage[T1]{fontenc}
\smartqed  

\RequirePackage{graphicx}
\RequirePackage{mathptmx}     
\usepackage{float}
\usepackage{tensor}
\usepackage{braket}
\usepackage{color}
\RequirePackage{flushend}
\RequirePackage[numbers,sort&compress]{natbib}
\usepackage[colorlinks,citecolor=blue,linkcolor=blue,anchorcolor=blue,filecolor=blue,
urlcolor=blue]{hyperref}
\usepackage{multirow}
\usepackage{graphicx}
\usepackage{tensor}
\usepackage[figuresright]{rotating}
\usepackage{booktabs}
\usepackage{hvfloat}
\usepackage{graphicx,color,xcolor}
\usepackage{amsfonts,amsmath}
\usepackage{amssymb,ulem}
\usepackage{epsfig,hyperref}
\usepackage{comment}
\RequirePackage{fix-cm}
\usepackage{siunitx}  
\usepackage{adjustbox}
\usepackage{amssymb}
\usepackage{pifont}
\newcommand{\xmark}{\text{\sffamily X}}%
\usepackage{orcidlink}

\journalname{Eur. Phys. J.} 
\begin{document}
	\title{Spherically symmetric anisotropic strange stars}
	\author{Luiz L. Lopes\thanksref{e1,addr1}
		\and
		H. C. Das\thanksref{e2, addr2}
	}
	\thankstext{e1}{e-mail: llopes@cefetmg.br}
	\thankstext{e2}{e-mail: harishchandra.das@ct.infn.it}
	\institute{Centro Federal de Educa\c{c}\~ao Tecnol\'ogica de Minas Gerais Campus VIII, Varginha/MG, CEP 37.022-560, Brazil \label{addr1}
		\and
		INFN Sezione di Catania, Dipartimento di Fisica, Via S. Sofia 64, 95123 Catania, Italy  \label{addr2}	
		 }
	\date{Received: xxxx / Accepted: xxxx}
	\maketitle
	\begin{abstract}
		In this work, we made an extensive study about the possible presence of anisotropies in strange stars. To accomplish this task, we use three different configurations for the strange matter: the unpaired matter, a two-flavor super-conducting (2SC) strange matter, and a fully three-flavor super-conducting strange matter (CFL). For each configuration, we calculate the relevant quantities for the strange stars, such as the mass-radius relation, the dimensionless tidal parameter, the moment of inertia, and the surface curvature for different degrees of anisotropies. Whenever possible, we compare our results with constraints found in the literature, especially focusing on the existence of very massive pulsars (PSR J0952-0607), as well as very light compact objects (HESS J1731-347).
\end{abstract}
\maketitle
\section{Introduction}
Strange stars are self-bounded compact objects composite by deconfined quarks. The theory of strange stars is based on the so-called Bodmer-Witten conjecture~\cite{Bodmer_1971, Witten_1984}. It assumes that the ordinary matter we know, composed of protons and neutrons, may only be meta-stable, while the true ground state of strongly interacting matter would therefore consist of the so-called strange matter (SM), which in turn is composed of deconfined up, down and strange quarks. For the SM hypothesis to be true, the energy per baryon of the deconfined phase (for $p = 0$ and $T = 0$) is lower than the nonstrange infinite baryonic matter, i.e., ($E_{\rm uds}/A < 930$ MeV), while at the same time, the nonstrange matter still needs to have an energy per baryon higher than the one of nonstrange infinite baryonic matter ($E_{\rm ud}/A > 930$ MeV); otherwise, protons and neutrons would decay into $u$ and $d$ quarks.

In this work, we discuss strange stars within three different possible scenarios. The normal, unpaired quark matter. Two flavors paired color-superconducting quark matter (2SC) and a full three flavors, color-conducting quark matter known as Color-Flavor-Locked (CFL) phase.  The quark matter is expected to behave in a full CFL phase at very high densities. However, at intermediate densities, the behavior of the quark matter is still an open issue~\cite{Alford2002, Alford2008}.

Furthermore, we study not only isotropic strange stars but also investigate the effects of local anisotropy. {The local anisotropy can arise due to exotic phenomena, such as a strong magnetic field, superfluity, phase transitions, etc. (see Ref.~\cite{Herrera_1997} for an extensive discussion). Several models have been proposed to explore the effects of pressure anisotropy inside the star \cite{Das_ILC_2022}. However, in this study, we use the so-called Bower-Liang (BL) model~\cite{Bowers_1974}}. In this simple model, the anisotropy is gravitationally induced and non-linear in pressure.  Moreover, the BL model has the advantage that the presence of local anisotropy does not break the spherical symmetry of the star, as discussed in Ref.~\cite{DelgadoeDelgado2018}.

We have already discussed anisotropic stars in our previous work~\cite{Lopes2023M}. However, in that short study, we considered only one model for the quark matter, the unpaired one. Moreover, that study also restricted our analysis to the mass-radius diagram and the tidal deformability parameter. In the present study, we extend our study in all aspects such as the unpaired matter and two models of color-superconducting phases. Moreover, altogether with the mass-radius relation and the dimensionless tidal parameter, we also discuss the effect of color-superconducting and anisotropy in the inertia moment of strange stars and the surface curvature~\cite{Kazim_2014}. This work is, therefore, complementary to other studies found in the literature.  For instance, the effect of anisotropy on radial oscillation was studied in Ref. \cite{Arbanil_2016}. The anisotropic charge in strange stars has been explored in the Tolman–Kuchowicz space-time geometry in  Ref.~\cite{Maurya_2019}. They got an interesting result that the pressure anisotropy initially dominates the Coulomb repulsive forces, and the repulsive forces dominate the anisotropic when the radius increases. On the other hand, in Ref.~\cite{Deb_2021}, the authors have explored the effect of the magnetic field on the anisotropic strange stars and observed that the transverse component of the magnetic field increases the mass and size of the anisotropic star and vice-versa for the radial component of the field.

We also check which of our results fulfill some constraints,  mainly focusing on those that are harder to explain for a traditional neutron star. The first is the HESS J1731-347 object~\cite{Doroshenko_2022} whose mass and radius are $M=0.77_{-0.17}^{+0.20}~M_\odot$ and $R = 10.4 _{-0.78}^{+0.86}$ km respectively. The true nature of the HESS object is one of the hot topics in nuclear astrophysics. In Ref.~\cite{Doroshenko_2022}, it was shown that some chiral theory can explain this object as an ordinary hadronic neutron star. The hadronic nature was also explored in Ref.~\cite{Kubis_2023}. On the other hand, the HESS object as a strange star was studied in refs.~\cite{Lopes2023M,Rather2023}. Finally, in Ref.~\cite{Sagun2023} the authors show that the
HESS object can be a hadronic neutron star with a soft EOS, or a hybrid star with an early deconfinement phase transition. The second constraint is the very small upper limit of the radius ($R_{1.4}<11.9$ km) of the canonical star,  given by Capano {\it et al.} ~\cite{Capano_2020}. Here again, we cannot completely rule out hadronic neutron stars, although such constraint is tight enough to rule out mostly hadronic EoSs. Finally, an extreme constraint is the {speculative}  mass of the black widow pulsars PSR J0952-0607, $M = 2.35\pm0.17 M_\odot$ \cite{Romani_2022}. Therefore, in this study, we provide an analysis of the anisotropic stars based on different models that could be able to fulfill all three strong constraints simultaneously.

Despite those extreme cases, we also investigate if our models are able to describe some more traditional constraints. For instance, the PSR J0740+6620 with a mass of $2.14 \ M_\odot$ was pointed in Ref.~\cite{Cromartie_2020}. Neutron Star Interior Composition Explorer (NICER) results refined this data, and today we have the PSR J0740+6620 with a mass of 2.08 $\pm$ 0.07 $M_\odot$ and a radius of 12.35 $\pm$ 0.35 km ~\cite{Miller_2021}. The canonical star, $M =1.4 M_\odot$, also received great attention in the last years. Two NICER results point that the radius of the canonical stars must be in the range $11.52 <R_{1.4}< 13.85$ km~\cite{Riley_2019}, and $11.96 <R_{1.4}< 14.26$ km~\cite{Miller_2019}. Nowadays, these results were revised to $11.80 <R_{1.4}< 13.10$ km ~\cite{Miller_2021}. We also investigate if our results are able to explain the mass-gap object in the GW190814 event~\cite{RAbbott_2020}, whose mass was estimated to be $2.50 - 2.67 \ M_\odot$. Still, there are several debates about its true nature; see Refs. 
~\cite{Lopes_ApJ, Das_PRD_2021}.

In addition, we also take the constraints from the gravitational wave observations by LIGO/VIRGO/KAGRA. The GW170817 event~\cite{Abbott_2017} put the constraints on the dimensionless tidal parameter of the canonical star  $\Lambda_{1.4}<800$~\cite{Abbott_2017}. This result was then refined in Ref.~\cite{Abbott_2018}, to  $70<\Lambda_{1.4}.<580$. Moreover, assuming that the mass-gap object in the GW190814 event was not a black hole implies that the dimensionless tidal parameter for the canonical star must be in the range of $458 <\Lambda_{1.4}<889$~\cite{RAbbott_2020}. 

Another important quantity is the moment of inertia (MOI) of the compact stars. Till now, we don't have any observational data for any NS. The authors of Ref.~\cite{Landry_2018} have obtained the MOI for several pulsars using the universal relations between the mass and the tidal deformability. Here, we calculate the MOI of the anisotropic SQS with different SQM models by varying the degrees of anisotropicity. One can constrain the value of MOI from the future observational data for the anisotropic compact star. There are other ways to constrain the magnitude of the MOI for different systems, such as Millisecond Pulsars (MSP), Double NS (DNS), and Low Mass X-ray Binary (LMXB), as done in Refs.~\cite{Landry_2018, Kumar_2019}. The MOI of these pulsars is expected to be measured soon; for example, the PSR J0737-3039(A) is the only known DNS up to date. Since the NS equation of state (EOS) is believed to be universal, the tidal deformability constraints from GW170817 have implications for all NSs, including PSR J0737-3039(A), the MOI has been obtained to be $1.15_{-0.24}^{+0.38}\times 10^{45}$ g cm$^2$ \cite{Landry_2018}. One can also estimate the MOI for other MSPs and LMXBs.

Finally, we discuss the surface curvature (SC) of the strange stars. The SC is a quantity used to measure the magnitude of the curvature made by a compact object due to its huge mass, and it is also one of the major consequences of Einstein's general theory of relativity. The SC is one of the observables that can be measured for different compact objects. Recently, the direct detection of gravity-field curvature has also been observed by using atom interferometers \cite{Rosi_2015}. In the future, we might have modern interferometers like LISA that could be able to answer the SC for compact objects such as neutron stars/white dwarfs. Since the SC mainly depends on macroscopic properties such as mass and radius, we can put constraints on the internal composition of the star. We explain the details of the methodology to obtain the SC in the following sections. 
\section{Quark matter models}
\subsection{Unpaired quark matter: the vector MIT bag model}
We begin our study with a normal, unpaired quark matter. This implies that all the flavors and colors are degenerated. Moreover, the particle distribution in the momentum space forms a sphere, the Fermi sphere, whose radius is the so-called Fermi momentum, $k_f$. At $T = 0$K, all particles are constrained inside the Fermi sphere. Consequently, the Fermi-Dirac distribution becomes the Heaviside step function~\cite{Greiner1995book}. To model the unpaired quark matter, we use a vector-enhanced MIT bag model~\cite{Klahn2015}.

The vector MIT bag model is an extension of the original MIT bag model~\cite{MITbag} that incorporates some features of the quantum hadrodynamics (QHD)~\cite{Serot_1992}. In its original form, the MIT bag model considers that each baryon is composed of three non-interacting quarks inside a bag. The bag, in its turn, corresponds to an infinity potential that confines the quarks. As a consequence, the quarks are free inside the bag and are forbidden to reach its exterior. All the information about the strong force relies on the bag pressure value, which mimics the vacuum pressure.

In the vector MIT bag model, the quarks are still confined inside the bag, but now they interact with each other through a vector meson exchange. This vector meson plays a role analog to the $\omega$ meson of the QHD~\cite{Serot_1992}. The Lagrangian density reads~\cite{Lopes_2021,Lopes_2021b}:
\begin{align}
\mathcal{L}_{\rm vMIT} &= \bigg\{ \bar{\psi}_q\big[\gamma^\mu(i\partial_\mu - g_{qV} V_\mu) - m_q\big]\psi_q  \nonumber \\
&
- B + \frac{1}{2}m_V^2V^\mu V_\mu  \bigg\}\Theta(\bar{\psi}_q\psi_q) ,
\label{vMIT}
\end{align}
where $m_q$ is the mass of the quark $q$ of ﬂavor $u$, $d$ or $s$, $\psi_q$ is the Dirac quark ﬁeld, $B$ is the constant vacuum pressure, and $\Theta(\bar{\psi}_q\psi_q)$ is the Heaviside step function to assure that the quarks exist only conﬁned to the bag. Imposing  mean-field approximation (MFA) and applying Euler-Lagrange to Eq.~(\ref{vMIT}), we obtain the energy eigenvalue for the quark, as well as the expected value for the vector field.
\begin{align}
E_q  = \mu_q =  \sqrt{m_q^2 + k^2} + g_{V}V_0, \label{EMT}
\end{align}
\begin{align}
m_V^2V_0 = \sum_q g_{V} n_q , \nonumber
\end{align}
where $n_q$ is the number density of the quark $q$ and $\mu_q$ is its chemical potential.
Now, applying  Fermi-Dirac statistics, the energy density is analogous to the QHD plus the bag term:
\begin{align}
\rho = \sum_q \frac{\gamma_q}{2\pi^2}\int_0^{k_f} dk \ k^2 \sqrt{k^2 + m_q^{2}}  +\frac{1}{2}m_V^2V_0^2  + B  \label{edMIT}
\end{align}
$\gamma_q=6~(3$ colors $\times \; 2$ spins$)$ is the degeneracy factor. Electrons are also added as a free Fermi gas to ensure chemical equilibrium. The pressure is then easily obtained by thermodynamics relations: $p =\mu n - \rho$. 

In this work we use $m_u = m_d = 4$ MeV, $m_s = 95$, and define $G_V~\equiv~(g_V/m_V)^2$, as suggested in Refs.~\cite{Lopes_2021,Lopes_2021b}. Although the bag constant $B$ and the parameter $G_V$ are not fully independent, they are weakly constrained to  only satisfy the stability window~\cite{Torres_2013}. This implies that the maximum mass of an isotropic strange star can vary from $1.61 \ M_\odot$ for $G_V = 0$~\cite{Lopes_2021}, up to $2.81 \ M_\odot$ for $G_V = 0.40$ fm$^2$~\cite{Lopes_ApJ}.

However, it is well accepted in the literature that color-superconducting matter is stiffer than unpaired one~\cite{Alford2008, Lugones2004, Ruster2004, Linares2006, Lorenzatto2022}.  We then use this fact to constrain the values of $G_V$ and $B$ by assuming that this model produces a softer EoS than the 2SC color-superconducting quark matter. Therefore, in this work, we take $G_V = 0.15$ fm$^2$ and $B^{1/4} = 150$ MeV.
\subsection{2SC color-superconducting quark matter}
The 2SC color-superconducting is the simplest pairing that evolves a BCS coupling, although it is also the least symmetrical one. In this model, the $s-$quark does not pair due to its high mass value. The lightest quarks form a Cooper pair if they present different colors and flavors (anti-symmetric coupling). As a consequence, only two of the three colors form Cooper pairs. Therefore, we have two of the three flavors and two of the three colors paring. This means that of the nine quarks, only four are paired~\cite {Ruster2004, Alford2008}. The presence of unpaired quarks gives rise to gap-less quasi-particles. The density of states of such quasi-particles is proportional to $\mu^2$ and, therefore, is very large~\cite{Shovkovy_2005}. Therefore, the 2SC EOS is not much stiffer than the unpaired one, as already suggested in Ref.~\cite{Ruster2004}.

In the 2SC color-superconducting matter, as in the unpaired one, a small quantity of electrons is needed to keep the electric charge net equal to zero. In Ref.~\cite{Agrawal2010}, the 2SC color-superconducting was explored within the NJL model. Moreover, in Ref.~\cite{Zdunik2013}, the authors show that the 2SC phase can be accurately described by an analytical approximation called constant-sound-speed (CSS) parametrization. The CSS model also seems to accurately reproduce the Field Correlator Method (FCM) with and without a color-superconducting phase~\cite{Alford2015, Burgio2016}.

In the CSS model, the pressure is a linear function of the energy density. Due to its simplicity and accuracy (see, for instance, Figs 1 and 2 in Ref.~\cite{Zdunik2013}), the CSS model is widely used in modern literature~\cite{Zdunik2013, Alford2013, Alford2015, Burgio2016, Han2019, Cierniak_2020, Sun2023, Lopes2023JCAP}. The CSS EoS and total number density read
\begin{align}
p &= a(\rho - \rho_{*}),  \nonumber  \\  \label{2SCEOS}
n &= n_{*}[(1 +a)p/(a\rho_{*})]^{1/(1+a)} .
\end{align}

We have, therefore, three free parameters: the square of the speed of sound ($v_s^2 = a$), the energy density at $p=0$ ($\rho_*$), which plays a role similar to the bag in the MIT bag models, and the number density at $p = 0$ ($n_*$),  which in turn, plays the role of the saturation density ($n_0$) of the MIT based models. In Ref.~\cite{Zdunik2013}, the authors freely vary the value of $a$ in the range $0.2 < a < 0.8$  and found that - depending on the NJL parametrization - the 2SC phase is well described with $a < 0.33$ while the CFL phase is described by $a > 0.35$.  Therefore, following that paper, we use $a = 0.302$, which was explicitly used in their fig. 3 to describe the 2SC quark matter.

We also use here $\rho_* = 205$ MeV/fm$^{3}$, and $n_* = 0.235$ fm$^{-3}$, which produce a energy per baryon $E/A = 890$ MeV, therefore, satisfying the Bodmer-Witten conjecture~\cite{Bodmer_1971,Witten_1984}.
\subsection{CFL color-superconducting quark matter}
If in the 2SC model, only four of the nine quarks are paired, in the CFL, all of them are paired in a fully anti-symmetric coupling in the space of colors and flavors~\cite{Alford2008}. Also, there are no gap-less quark quasi-particles in the low-energy spectrum of the CFL phase due to the color Meissner effect~\cite{Shovkovy_2005}. Moreover, the Cooper pairing in the CFL phase helps to enforce the equal number densities of all three quark flavors, $n_u = n_d = n_s$, implying that electrons are not present in this phase. In general, the CFL phase is expected at densities above four times the nuclear saturation density, $n_0$, while the 2SC phase is expected in the region $2~<n/n_0~<4$~\cite{Zdunik2013}. However, some studies indicate that the 2SC phase is not favored at all due to the significant free energy cost of the 2SC phase~\cite{Alford2002}. In this case, the unpaired quark matter undergoes a phase transition to a full CFL phase, or even a more exotic phase, as a non-BCS pattern. The gap parameter ($\Delta$) that determines the pairing strength of Cooper pairs can be very high in the CFL phase, above 250 MeV~\cite{Lugones2004}. In this case, a very stiff EoS and consequently a very massive strange star is produced.

As we did in the 2SC phase, we also use the CSS model (Eq.~\ref{2SCEOS}) as an analytical approximation to the CFL phase. Here, following the Ref.~\cite{Zdunik2013}, we use $a =0.57$, as it was explicitly used in their fig. 3 to the CFL phase. We also use $\rho_*$ = 209 MeV/fm$^3$, and $n_*$ = 0.235 fm$^{-3}$, which produce a energy per baryon $E/A = 897$ MeV, satisfying the Bodmer-Witten conjecture.
\section{Spherically symmetric anisotropic stars}
\subsection{Hydrostatic equilibrium}

\begin{figure}
    \centering
    \includegraphics[width=0.45\textwidth]{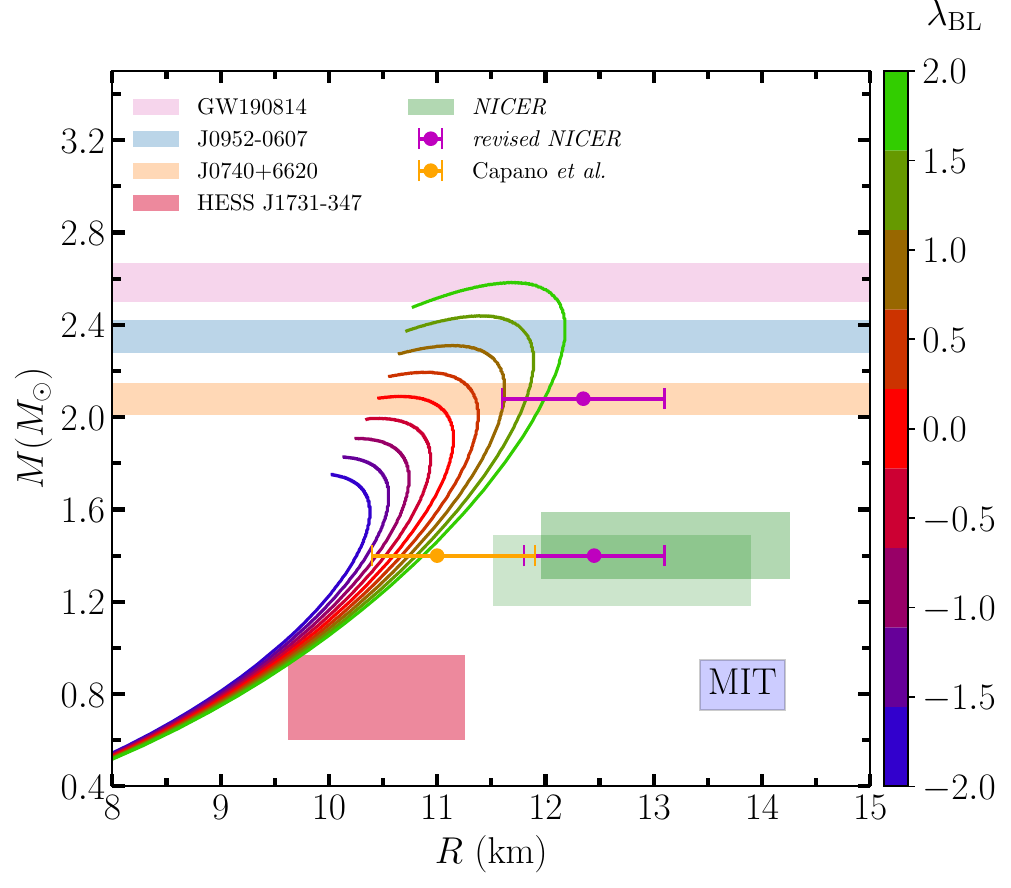} \\
     \includegraphics[width=0.45\textwidth]{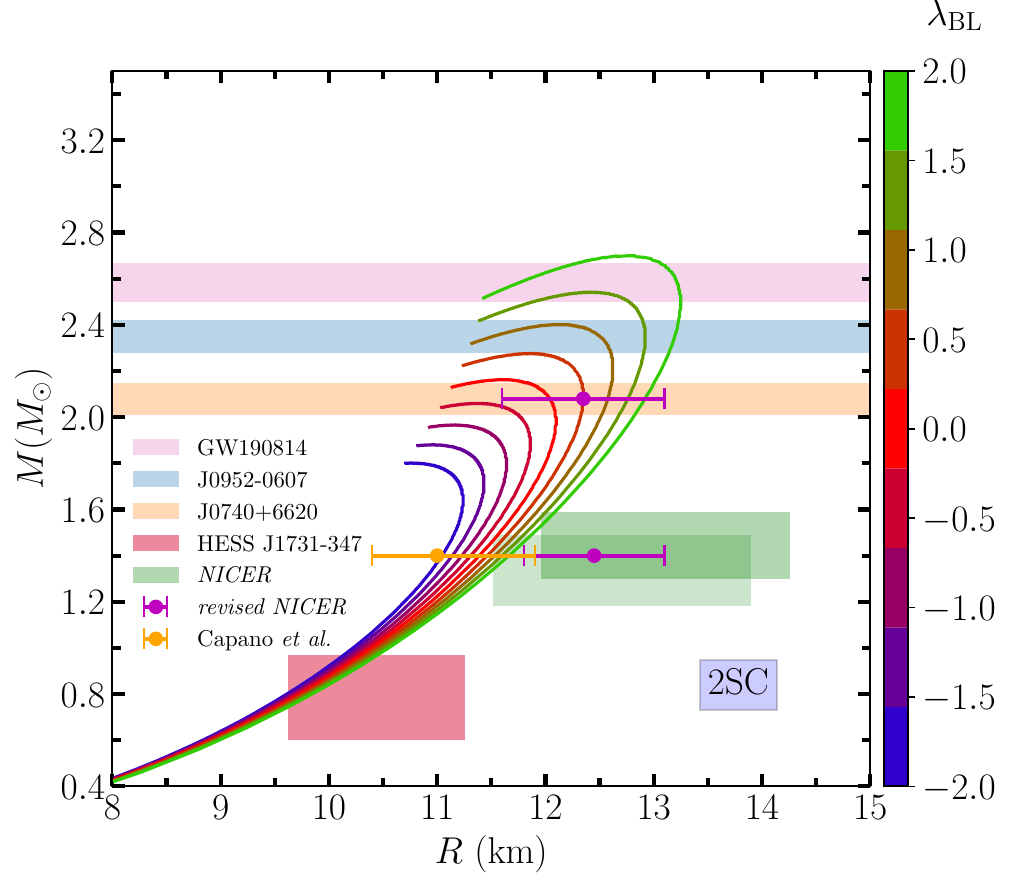} \\
       \includegraphics[width=0.45\textwidth]{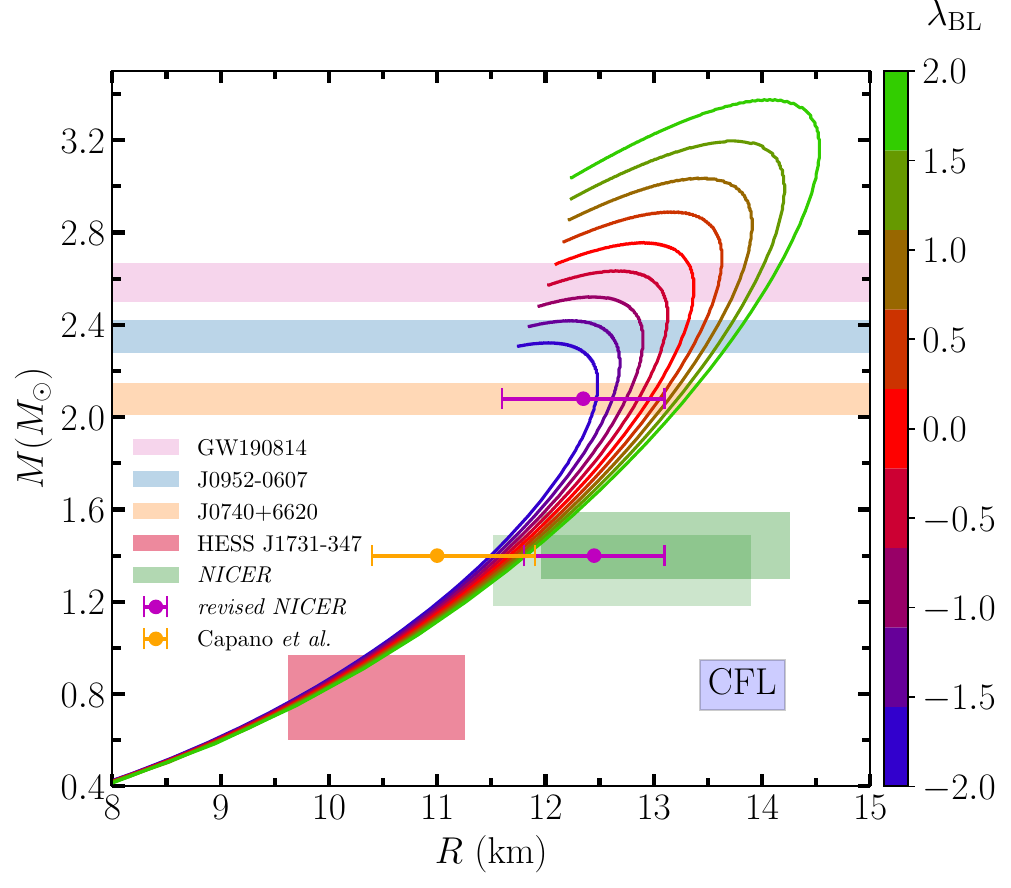}
    \caption{Mass-radius relation for unpaired (top), 2SC (middle), and  CFL (bottom) strange stars for different values of anisotropies. The hatched areas are bounds discussed in the text.}
    \label{tovs}
\end{figure}

In this section, we briefly review the theory of static equilibrium distribution of matter, which is spherically symmetric but whose stress tensor is, in general, locally anisotropic, as originally introduced in Ref.~\cite{Bowers_1974}. In the isotropic case, the stress-energy tensor reads: $T_{\mu\nu}$ = diag($\rho, -p,-p, -p$). We now introduce anisotropies without breaking the spherical symmetry by assuming the following stress-energy tensor: $T_{\mu\nu}$ = diag($\rho, -p_r,-p_t, -p_t$). Spherical symmetry implies that (in canonical coordinates) the stress-energy tensor $T_{\mu\nu}$ is diagonal, and moreover $p_\phi =p_\theta = p_t$~\cite{DelgadoeDelgado2018}.

 We now can redefine the anisotropic energy-momentum tensor as done in Refs. ~\cite{Doneva_2012, Silva_2015, Estevez_2018}:
\begin{align}
    T_{\mu\nu} = (\rho+p_t)u_\mu u_\nu + (p_r-p_t) k_\mu k_\nu + p_t g_{\mu\nu},
    \label{eq:tmunu_aniso}
\end{align}
where $\rho$, $p_r$, and $p_t$ are the energy density, radial pressure, and tangential pressure, respectively. $k_\mu$ is the unit radial vector ($k^\mu k_\mu = 1$) with $u^\mu k_\mu = 0$. The Schwarzschild metric for this type of star having the spherically symmetric and static configuration is defined as 
\begin{align}
    ds^2= e^{\nu}dt^2-e^{\lambda}dr^2-r^2d\theta^2-r^2 \sin^2\theta  d\varphi^2\,, \label{Scz}
\end{align} 
where $r$, $\theta$, and $\phi$ are the Schwarzschild coordinates. 

Applying Einstein's field equations, we obtain:

\begin{align}
\frac{\lambda'}{r}e^{\lambda} + \frac{1}{r^2}(1 - e^{\lambda}) = 8\pi\rho , \label{elambda}
\end{align}

\begin{align}
\frac{\nu'}{r}e^{\lambda} - \frac{1}{r^2}(1 - e^{\lambda}) = 8\pi p_r , \label{enu}
\end{align}
and the contracted Bianchi identities give us:

\begin{align}
\frac{dp_r}{dr} = - (\rho + p_r)\frac{\nu'}{2} + \frac{2}{r}(p_t - p_r). \label{edp}
\end{align}

Finally, by isolating the $\nu'$ in Eq.~\ref{enu} and replacing it in Eq.~\ref{edp}, we  can write
the equilibrium equations in the Tolman-Oppenheimer-Volkoff form~\cite{Doneva_2012}:

\begin{align}
    \frac{dp_r}{dr}=-\frac{\left( \rho + p_r \right)\left(m + 4\pi r^3 p_r \right)}{r\left(r -2m\right)} +\frac{2\sigma}{r} \,,
    \label{tov1:eps}
\end{align}
\begin{align}
    \frac{dm}{dr}=4\pi r^{2}\rho\,,
    \label{tov2:eps}
\end{align}
where $\sigma=p_t-p_r$ is the anisotropy parameter, `$m$' is the mass enclosed within the radius $r$. The radial pressure is then obtained from a pre-determined EOS.  On the other hand, for the case of the transverse pressure, we use the BL model in the following \cite{Bowers_1974}:
\begin{align}
    \label{Anisotropy_eos}
    p_t = p_r + \frac{\lambda_{\rm BL}}{3} \frac{(\rho+3p_r)(\rho + p_r)r^2}{1-2m/r} \,,
\end{align}
where the factor $\lambda_{\rm BL}$ measures the degree of anisotropy in the fluid. There are some boundary conditions required to solve the above Eqs. (\ref{tov1:eps}-\ref{Anisotropy_eos}) as done in Refs. \cite{Biswas_2019, Das_ILC_2022}. Also, different fluid conditions must be satisfied for the anisotropic stars such as (i) $p_r, p_t$, and $\rho > 0$, (ii) $0<c_{\rm s, t}^2<1$, (iii)  $p_r = p_t$ for $r =0$, etc. Other conditions are mentioned in Ref.~\cite{Das_ILC_2022}. We plot the mass-radius relation for different values of $\lambda_{\rm BL}$ for unpaired, 2SC, and CFL strange stars in Figure \ref{tovs}.

We begin our analysis by comparing the different models of quark matter in the symmetric case.  It is observed that the unpaired matter produces a maximum mass of $2.09 \ M_\odot$,  the 2SC produces a maximum mass of $2.16 \ M_\odot$, and the CFL produces an impressive maximum mass of $2.76 \ M_\odot$.

In the case of the three main constraints, we see that the unpaired matter is not able to reproduce the radius of the HESS object~\cite{Doroshenko_2022} in the absence of anisotropies, while the 2SC and CFL easily fulfill this task. In relation to the constraint related to the radius of the canonical star $10.4<R_{1.4}<11.9$ km, presented in Ref.~\cite{Capano_2020}, all quark models are able to fulfill it in the isotropic case. Our last main constraint is the speculative mass of the black widow pulsars PSR J0952-0607,  $M = 2.35\pm0.17 M_\odot$ \cite{Romani_2022}. In this case, only the CFL quark matter is able to reach such high mass, although the 2SC is very close. 

In relation to the other constraints, we see that no isotropic model is able to fulfill the revised NICER results for the canonical star, $11.80 <R_{1.4}< 13.10$ km ~\cite{Miller_2021} (Notice that the overlap between the results in Ref.~\cite{Capano_2020} and those in Ref.~\cite{Miller_2021} is only 100 m), but the CFL still fulfill its original range ~\cite{Riley_2019}. In relation to the  PSR J0740+6620~\cite{Miller_2021}, we see that all quark models reach the mass range  of 2.08 $\pm$ 0.07 $M_\odot$, but the unpaired matter produces  too-low radii, in disagreement with the range of 12.35 $\pm$ 0.35 km. Finally, the mass-gap object in the GW190814 event, with $M~>2.50~M_\odot$~\cite{RAbbott_2020}, can be explained only by the CFL phase. Before we finish the analysis of the isotropic case, it is worth pointing out that most of the constraint can be fulfilled with unpaired matter by simply increasing the value of $G_V$~\cite{Lopes_ApJ}. However, in this work, we impose that unpaired matter must be softer than the super-conducting one. 

We now analyze the effects of anisotropies. As expected, a positive value of $\lambda_{\rm BL}$ increases the maximum mass, as well as increases the radius of a star with fixed mass~\cite{Bowers_1974, Lopes2023M}, while negative values do the opposite. For the extreme positive value ($\lambda_{\rm BL} = +2.0$), we see that the maximum mass increases by around 23$\%$ in the MIT model, 25$\%$ in the 2SC, and 22$\%$ in the CFL phase.  In the extreme negative value ($\lambda_{\rm BL} = -2.0$), we see that the maximum mass decreases by around 16$\%$ in all three quark models.

In relation to the radius of the canonical star, the extreme positive value ($\lambda_{\rm BL} = +2.0$) causes an increase by around 2.5$\%$ in the MIT model, 2.3$\%$ in the 2SC, and 1.0$\%$ in the CFL phase. In the extreme negative value ($\lambda_{\rm BL} = -2.0$), we see a decrease of around 3.2$\%$ in the unpaired matter, 3.0$\%$ in the 2SC, and 1.8$\%$ in the CFL phase. 

We can conclude that the effect of the anisotropies affects more the maximum mass than the radius of the canonical star. At the same time, its effects in the maximum mass are almost independent of the quark matter EOS, but the effects on the canonical radius are strong for a softer EOS.

In relation to the constraints, we see that a positive value of $\lambda_{\rm BL}$ will improve the results of the MIT. For instance, for $\lambda_{\rm BL}~>0.5$, the HESS object can be described even in the unpaired quark matter. For $\lambda_{\rm BL}~>1.0$, the radius range of the PSR J0740+6620~\cite{Miller_2021} is satisfied, as well as the mass of the black widow pulsars PSR J0952-0607. Additionally, for $\lambda_{\rm BL} = 2.0$, even the mass-gap object in the GW190814 event can be explained as an unpaired quark matter. For the 2SC, we see that  $\lambda_{\rm BL}~>0.5$ is enough to fulfill the mass range of the black widow pulsars PSR J0952-0607, while for $\lambda_{\rm BL} = 1.5$, the mass of the secondary object in the GW190814 event is reached. Finally, in the case of the CFL phase, we have a different behavior. As the CFL matter presents a stiff EOS even in the isotropic case, if we increase $\lambda_{\rm BL}$ up to values of 1.5, their curves do not cross the radius range of the PSR J0740+6620 pulsar~\cite{Miller_2021}. On the other hand, most of the constraints are satisfied even for extremely negative values of $\lambda_{\rm BL}$.
\subsection{Tidal deformation}
The star shape is deformed due to its presence in the external field of its companion star. The degree of deformation is measured by the parameter $\lambda$, which is defined as the tidal deformability of a star. The definition for the dimensionless tidal deformability $\Lambda$, which is a quantity observed by the LIGO/Virgo, and it has a unique relation with the tidal Love number ($k_2$), and the compactness $(C)$ of the star is~\cite{Hinderer_2008, Hinderer_2009}: 
\begin{align}
 \Lambda = \frac{2}{3} k_2 C^{-5}, ~\label{tidaleq}
\end{align}
\begin{figure}
    \centering
    \includegraphics[width=0.45\textwidth]{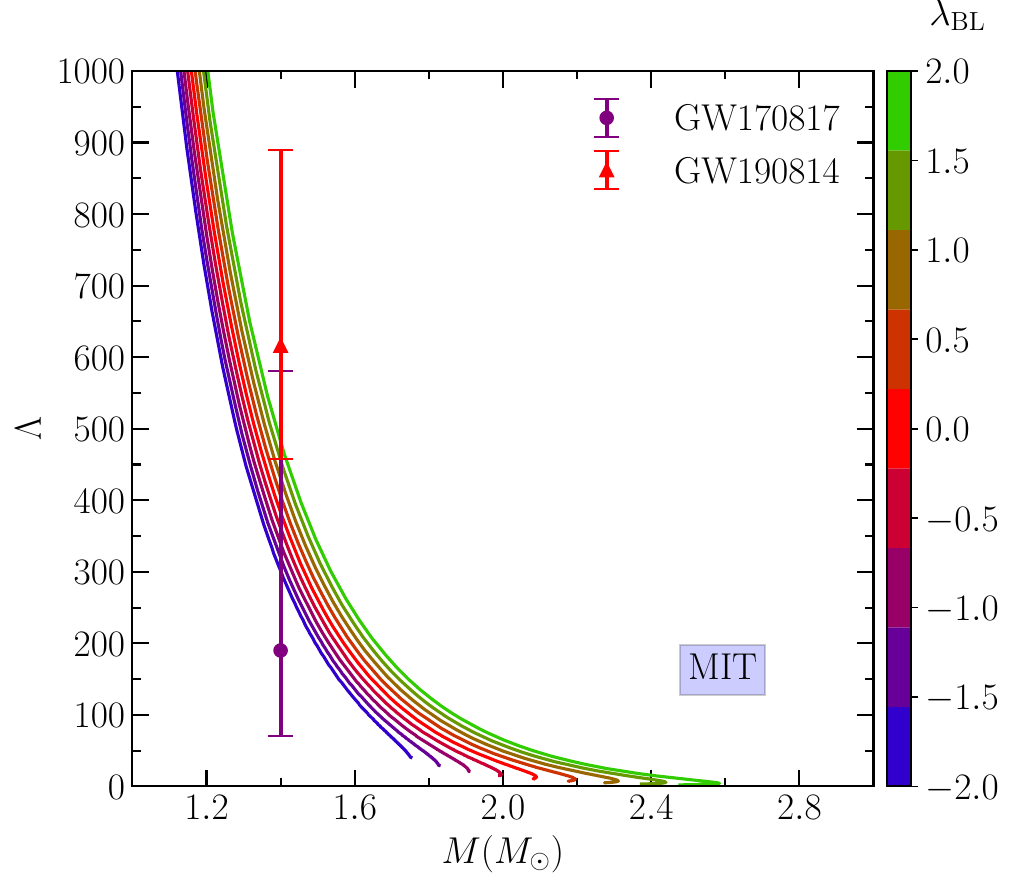} \\
     \includegraphics[width=0.45\textwidth]{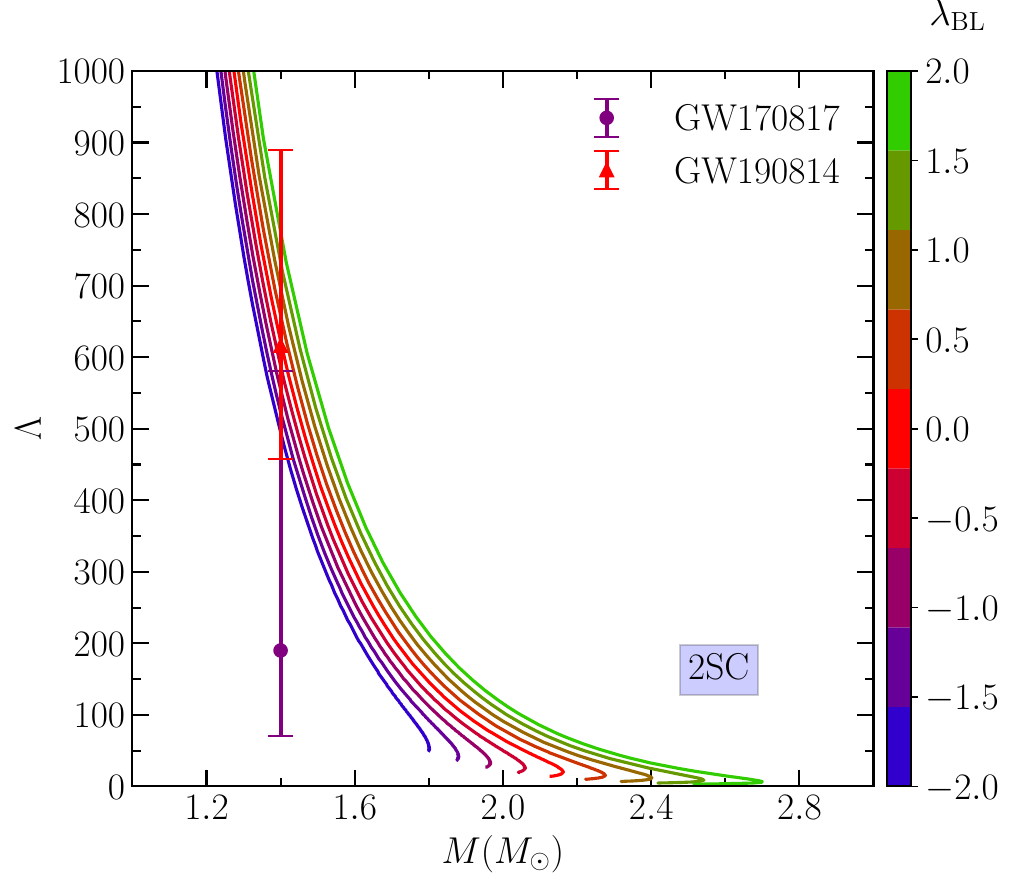} \\
       \includegraphics[width=0.45\textwidth]{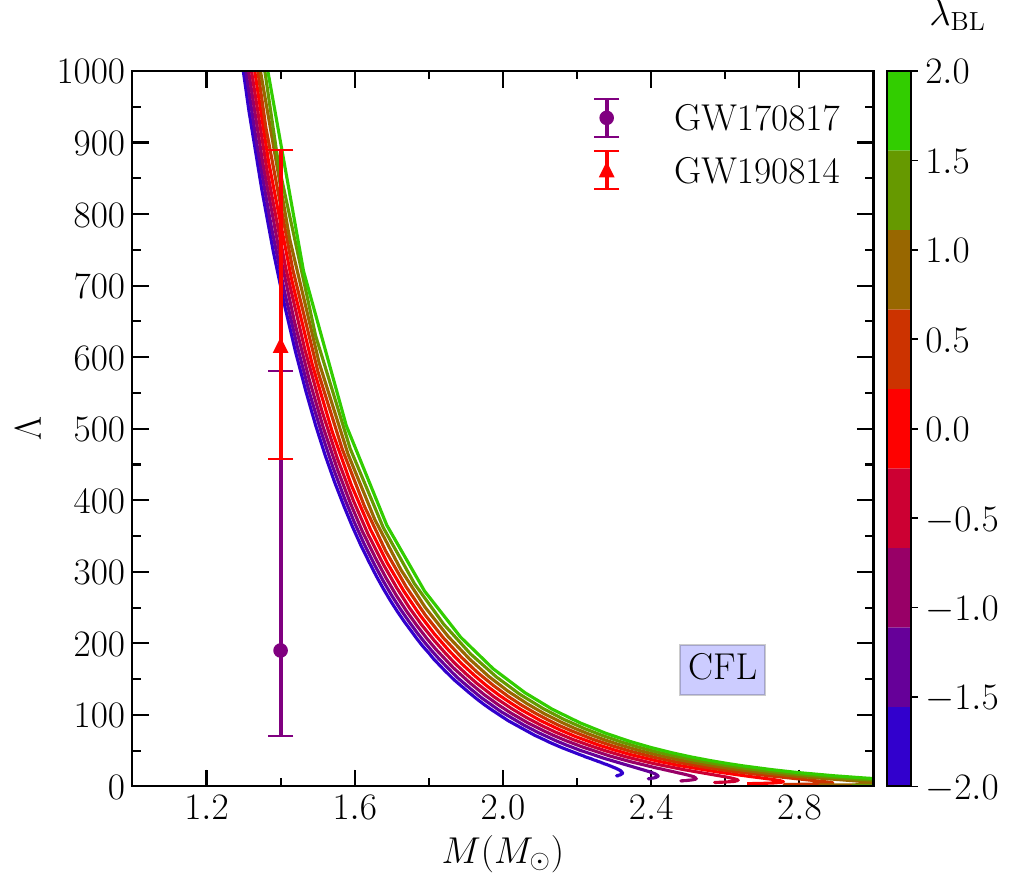}
\caption{Dimensionless tidal parameter ($\Lambda$) for unpaired (top), 2SC (middle), and  CFL (bottom) strange stars for different values of anisotropies. The bounds related to the GW170817 and GW190814 events are also shown.}
\label{tidal}
\end{figure}
We use the linear perturbation in the Throne and Campolattaro metric to determine the value of $k_2$, \cite{Throne_1967}. In the case of anisotropic stars, we use the formalism originally introduced in Ref.~\cite{Biswas_2019}:
\begin{align}
&H^{''} + H^{'} \bigg[\frac{2}{r} + e^{\lambda} \left(\frac{2m(r)}{r^2} + 4 \pi r (p_r - {\rho})\right)\bigg] 
\nonumber \\
&+ H \left[4\pi e^{\lambda} \left(4\rho + 8p_r + \frac{\rho + p_r}{dp_t/d\rho}(1+c_s^2)\right) -\frac{6 e^{\lambda}}{r^2} - {\nu^\prime}^2\right] 
= 0\,.
\end{align}
The term $dp_t/d\rho$ represents the change of $p_t$ with respect to energy density for a fixed value of $\lambda_{\rm BL}$. The internal and external solutions to the perturbed variable $H$ at the star's surface can be matched to get the tidal Love number \cite{Damour_2009, Hinderer_2008}. The value of the tidal Love number can then be calculated using the equation in the following ~\cite{Hinderer_2008, Hinderer_2009, Flores_2020, DasBig_2021}.  
\begin{align}
    k_2 &= \, \frac{8}{5} C^5 (1-2C)^2 \big[ 2(y_2-1)C - y_2 + 2 \big]
    \nonumber \\ 
    &\times \Big\{ 2C \big[ 4(y_2+1)C^4 + 2(3y_2-2)C^3 - 2(11y_2-13)C^2 
    \nonumber \\ &
    + 3(5y_2-8)C - 3(y_2-2) \big]+ 3(1-2C)^2 
    \nonumber \\ &
    \times \big[ 2(y_2-1)C-y_2+2 \big] \log(1-2C) \Big\}^{-1} \, , 
    \label{eq:k2}
\end{align}
where $y_2$ depends on the surface value of $H$ and its derivative 
\begin{equation}
    y_2 = \frac{rH^{'}}{H}\Big|_R - \frac{4\pi R^3 \rho_s}{M},
\end{equation}
where the $\rho_s$ is the energy difference between the internal and external regions. The results are presented in Figure~\ref{tidal}.

As in the last sub-section, we begin our analyses with the isotropic case. In the case of unpaired matter, it provides the softer EOS, which predicts the lower values of $\Lambda$. For the canonical mass, we have $\Lambda_{1.4} = 383$. As we consider color-superconducting, the EOS becomes stiffer, and the dimensionless tidal parameter increases. The values of  $\Lambda_{1.4} = 622$ and $787$ for the 2SC and the CFL phases, respectively. In relation to the effects of the anisotropies on the dimensionless tidal parameter, we see that the higher values of $\lambda_{\rm BL}$ predict the higher value of $\Lambda_{1.4}$ and vice-versa. For an extreme positive value, $\lambda_{\rm BL} = +2.0$, we see an increase of the $\Lambda_{1.4}$ by around $26\%$ in the MIT model, $24\%$ in the 2SC, and $14\%$ in the CFL phase.  For an extreme negative value, $\lambda_{\rm BL} = -2.0$, it is observed that the decrease of the $\Lambda_{1.4}$ by around $22\%$ in the MIT model, $20\%$ in the 2SC, and $11\%$ in the CFL phase. Therefore, as in the case of the radius of the canonical mass, the effects of anisotropies in the dimensionless tidal parameter are stronger in softer EOS.

We now discuss the results in the light of two constraints: the GW170817 ($70<\Lambda_{1.4}.<580$)~\cite{Abbott_2018} and the  GW190814 event ($458 <\Lambda_{1.4}< 889$)~\cite{RAbbott_2020}, although it is worth to emphasize that the nature of the mass-gap object in the GW190814 is not know yet, therefore such constraint is still speculative. We see that the MIT model fulfills the GW170817 constraint for all values of $\lambda_{\rm BL}$. Moreover, for $\lambda_{\rm BL}$ = +2.0, the constraint of GW190814 is also satisfied.  For the 2SC, we see that the GW170817 is satisfied for values up to $\lambda_{\rm BL} = -1.0$, and the GW190814 for all degrees of anisotropies. Finally, the CFL model fails to fulfill the GW170817 for all values of $\lambda_{\rm BL}$, but it can describe the GW190814 limit if $\lambda_{\rm BL}~<+2.0$.
\subsection{Moment of Inertia}
\label{moi}
For a slowly rotating NS, the equilibrium position can be obtained by solving Einstein's equation in the Hartle-Throne metric, as given in Refs.~\cite{Hartle_1967, Hartle_1968, Hartle_1973}:
\begin{align}
    ds^2  = & -e^{2\nu} \ dt^2 + e^{2\lambda} \ dr + r^2 \ (d\theta^2 +\sin^2\theta d\phi^2)
    \\ \nonumber
    & - 2\omega(r)r^2\sin^2\theta \ dt \ d\phi
\end{align}
\begin{figure}
    \centering
    \includegraphics[width=0.45\textwidth]{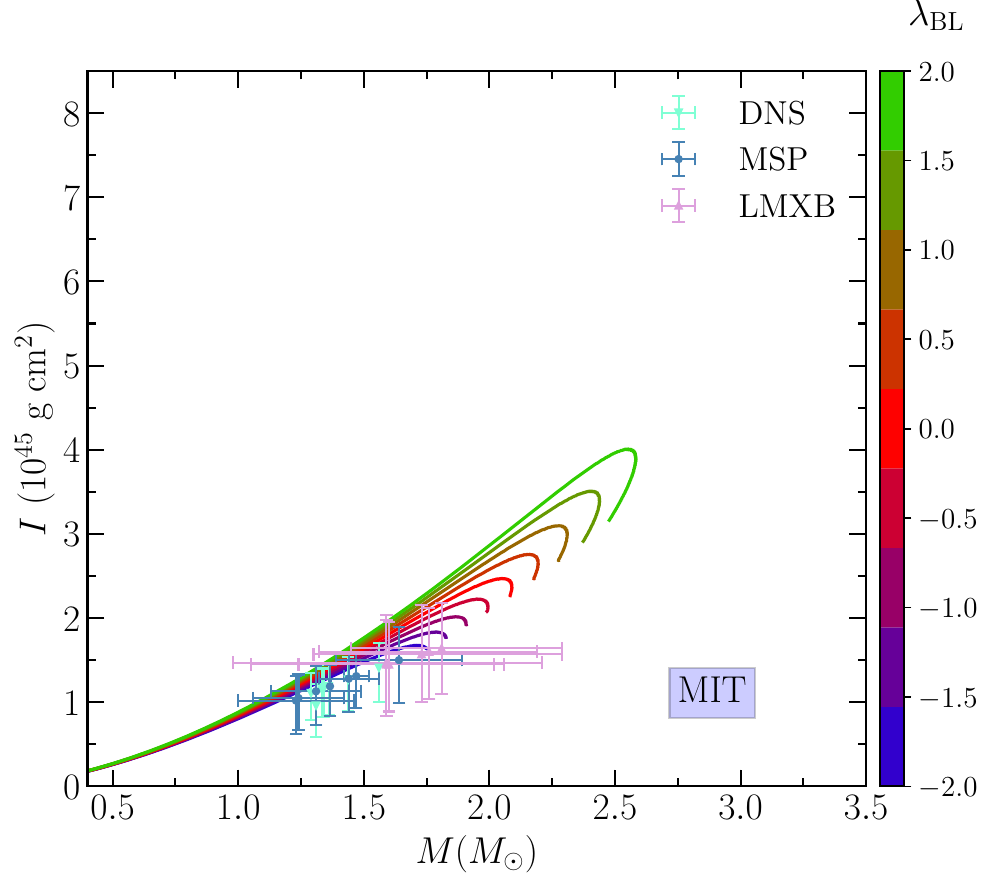} \\
     \includegraphics[width=0.45\textwidth]{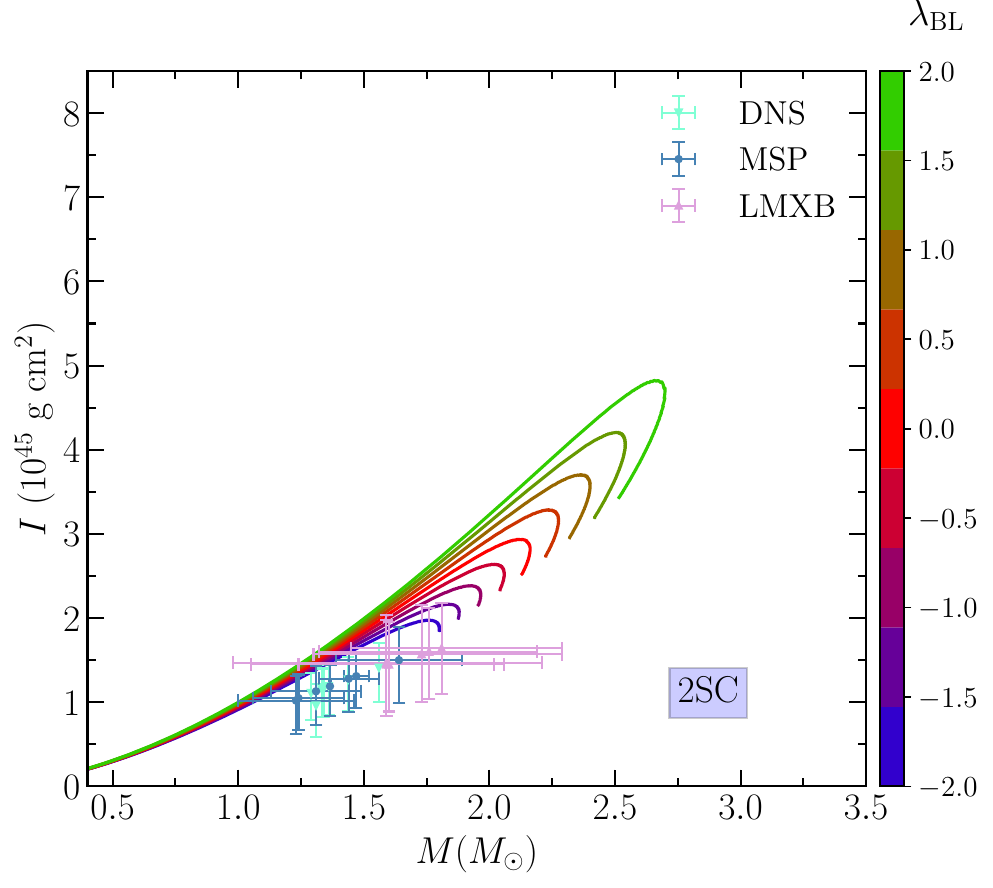} \\
       \includegraphics[width=0.45\textwidth]{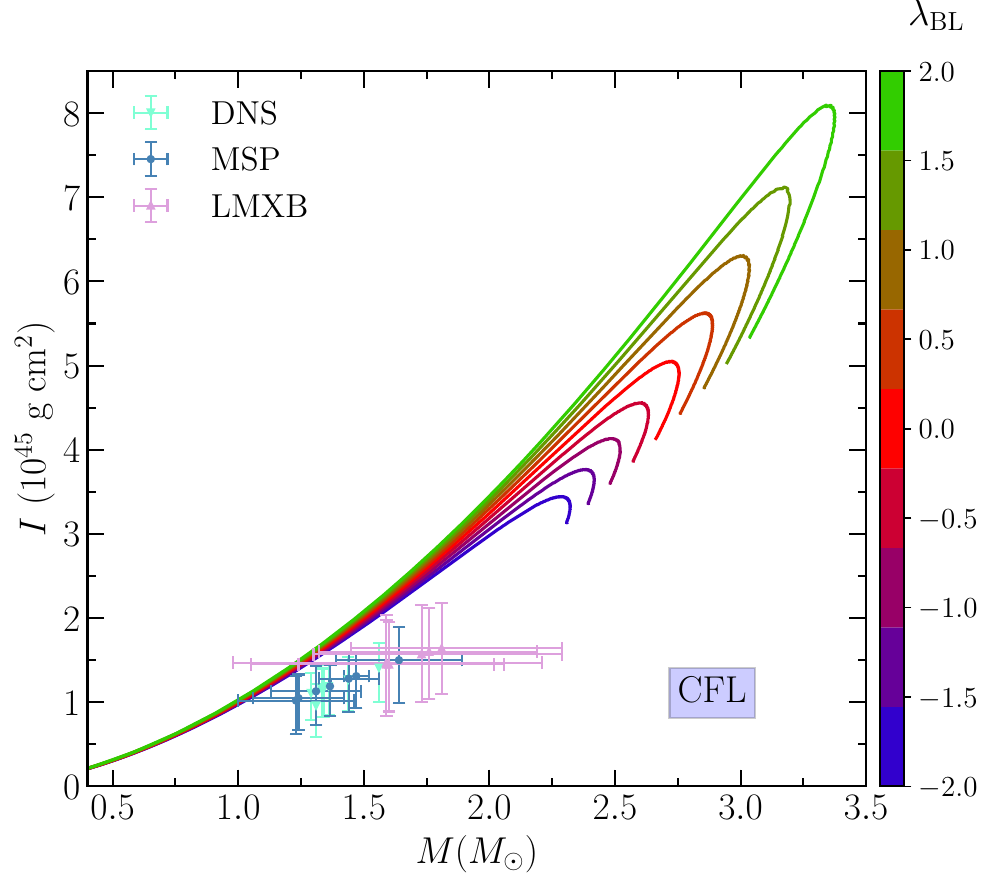}
    \caption{Moment of Inertia ($I$) for unpaired (top), 2SC (middle), and  CFL (bottom) strange stars for different values of anisotropies. The bounds are discussed in the text and in Refs.~\cite{Landry_2018, Kumar_2019}.}
    \label{MOI}
\end{figure}
For a slowly rotating anisotropic compact star, the MOI can be  calculated as in Ref.~\cite{Sulaksono_2020}:
\begin{align}
    I = \frac{8\pi}{3}\int_0^R \frac{r^5J\tilde{\omega}}{r- 2M}(\rho+p_r)\left[1+\frac{\sigma}{\rho+p_r}\right] \, dr,
    \label{eq:MI}
\end{align}
where,
\begin{align}
 \sigma = p_t - p_r = \frac{\lambda_{\rm BL}}{3} \frac{(\rho+3p_r)(\rho + p_r)r^2}{1-2m/r}, \label{sigmapp} 
\end{align}
and $\tilde{\omega}=\Bar{\omega}/\Omega$, where $\Bar{\omega}$ is the frame dragging angular frequency, $\Bar{\omega} = \Omega-\omega(r)$. $J$ is defined as $e^{-\nu}(1-2m/r)^{1/2}$.  Hence Eq. (\ref{eq:MI}), can be rewritten using Eq. (\ref{Anisotropy_eos}) as
\begin{align}
    I &= \frac{8\pi}{3}\int_0^R \frac{r^5J\Tilde{\omega}}{r- 2M}(\rho+p_r)\left[1+\frac{\frac{\lambda_{\rm BL}}{3} \frac{(\rho+3p_r)(\rho + p_r)r^2}{1-2m/r}}{\rho+p_r}\right] \, dr,
\end{align}

As already pointed out in Ref.~\cite{Sulaksono_2020}, a measurement of NS moment inertia is crucial because its inherent capability to constrain quite restricts the EOS of NSs at high density are insensitive to EOS, and it has a universal relation with compactness and tidal deformability. The results are displayed in Fig.~\ref{MOI}.

At first approximation, the MOI is proportional to $MR^2$. Therefore, the larger and heavier strange stars will present a significantly higher MOI than smaller and lighter ones. This is reflected in Fig.~\ref{MOI}. Considering the isotropic case, we see that the MOI for the canonical star ($I_{1.4}~\times~10^{45}$~g.cm$^{-2}$) are 1.47 for the unpaired matter, 1.66, 1.78 for the 2SC and the CFL respectively. In relation to the maximally massive strange stars, the MOI ($I_{\rm max}~\times~10^{45}$~g cm$^{-2}$) are 2.36, 2.82, and 4.92 for MIT, 2SC, and CFL, respectively. We can also see that the MOI is strongly affected by anisotropies. For positive values of $\lambda_{\rm BL}$ the increase of the $I_{1.4}$ can reach $6.8\%$ in the MIT, $6.6\%$ for the 2SC, and 3.4$\%$ for the CFL. For negative values of $\lambda_{\rm BL}$, we have a decrease of the $I_{1.4}$ of 8.1$\%$ in the MIT, $7.8\%$ in the 2SC, and $4.5\%$ for the CFL. These results are expected; as pointed out earlier, the effects of the anisotropy in the canonical star are stronger in the softer EoS. On the other hand, in relation to the maximum mass of strange stars, we see for positive values of $\lambda_{\rm BL}$, an increase of 65 $\%$ in the MIT, 67$\%$ in the 2SC, and 62$\%$ in the CFL. For negative values of $\lambda_{\rm BL}$, we have a decrease of $29\%$ in the MIT, $33\%$ in the 2SC, and $32\%$ in the CFL. These large variations reflect the dependence of the MOI with $MR^2$.

In relation to the constraints, due to the high uncertainty in the mass, most of the results satisfy the DNS, MPS, and LMXB bounds discussed in Refs.~\cite{Landry_2018, Kumar_2019}. However, for higher values of $\lambda_{\rm BL}$ in the 2SC case, some curves do not cross some of the DNS and MSP regions. In the CFL case, even the isotropic case does not cross some curves of the MSP bound. Nevertheless, we pay special attention to the constraint of the PSR J0737-3039(A). With a well-established mass of 1.34 $M_\odot$, in Ref.~\cite{Landry_2018} was able to constraint its MOI to the range $1.15_{-0.24}^{+0.38}\times 10^{45}$ g cm$^2$.  For the MIT, we see that such constraint is satisfied by all values of $\lambda_{BL}$. For the 2SC it is satisfied if $\lambda_{BL}~<+1.5$, and for the CFL, it is satisfied only if  $\lambda_{\rm BL}~< -0.5$.
\begin{figure}
    \centering
    \includegraphics[width=0.45\textwidth]{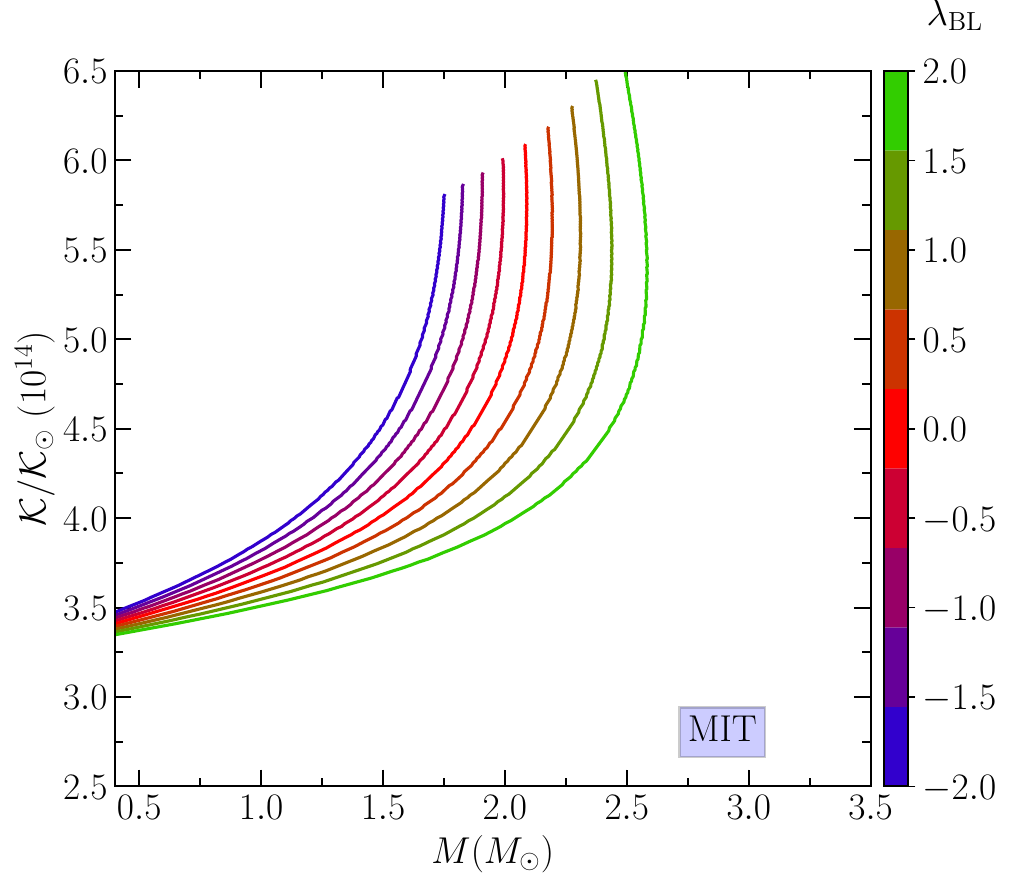} \\
     \includegraphics[width=0.45\textwidth]{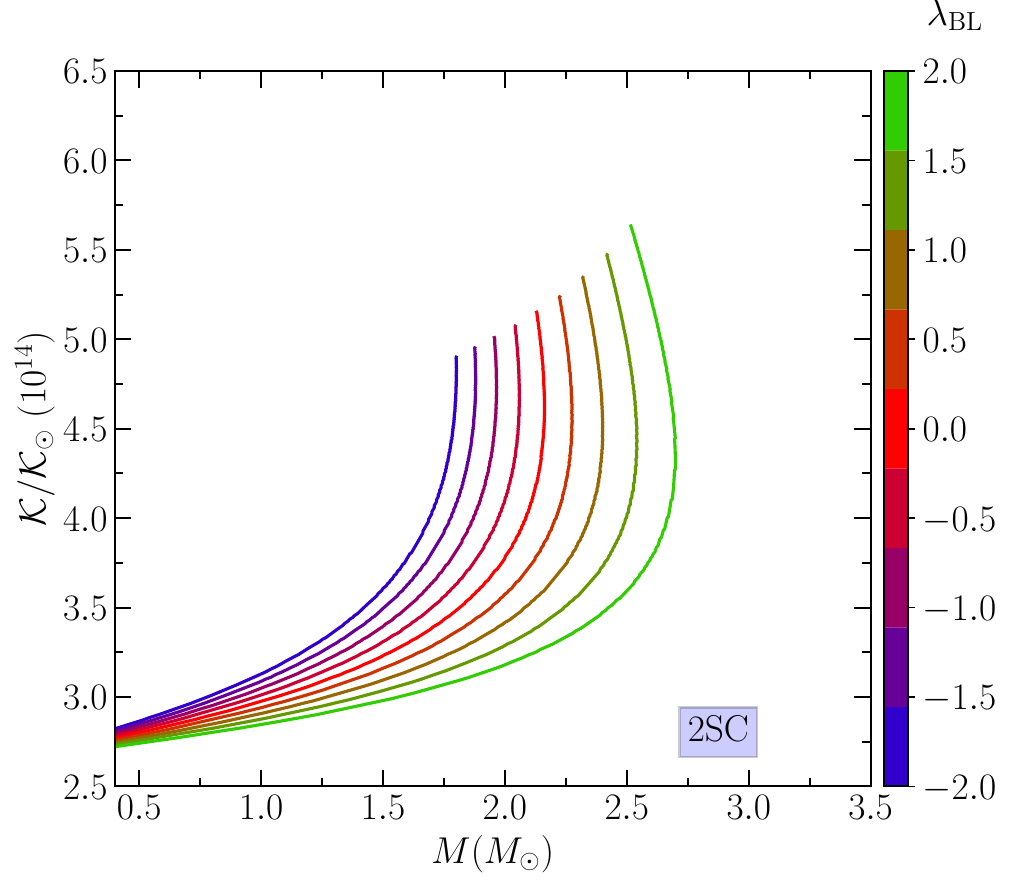} \\
       \includegraphics[width=0.45\textwidth]{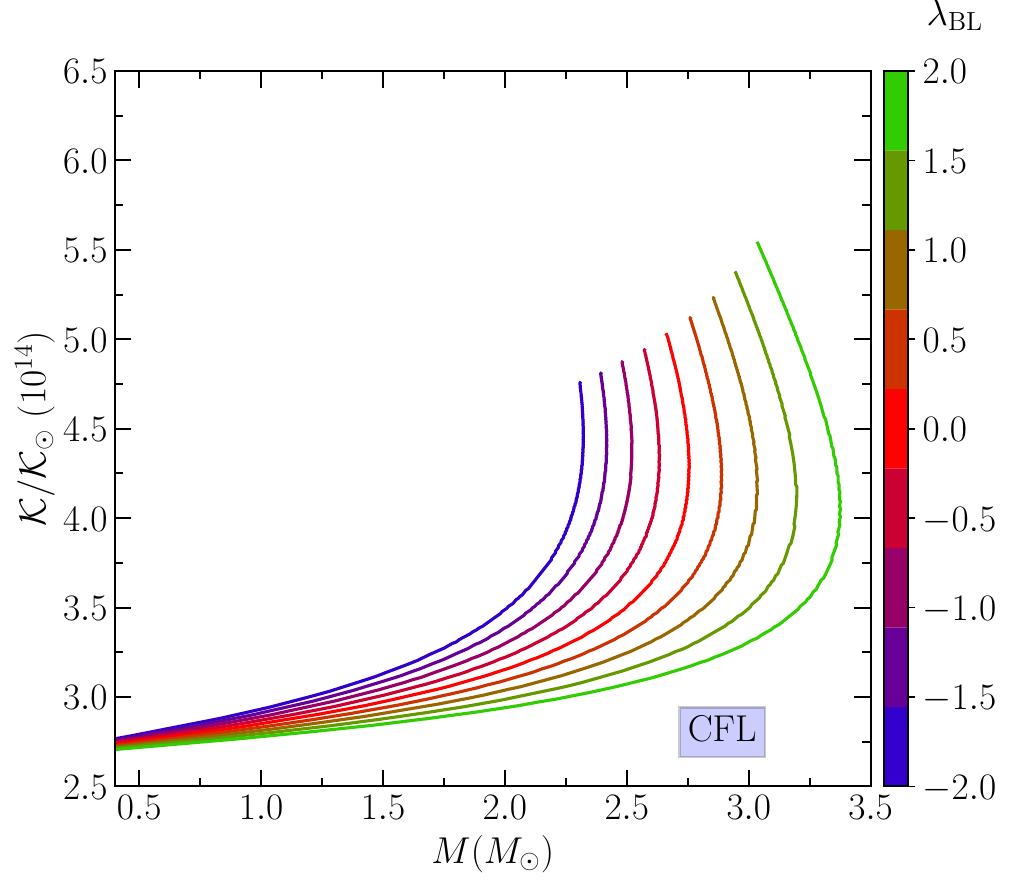}
    \caption{ Surface curvature (SC) for unpaired (top), 2SC (middle), and  CFL (bottom) strange stars for different values of anisotropies.}
    \label{SC}
\end{figure}
\subsection{Surface Curvature}
\label{sc}
In the general theory of relativity (GR), there are different quantities to measure the curvature of space-time, mainly the Ricci scalar, the Ricci tensor, the Weyl tensor, and the Riemann tensor. However, except for the Riemann tensor, all of the magnitude becomes zero outside the star. Therefore, we adopt the curvature quantity from the Refs. \cite{Kazim_2014, Das_2021}, that is known as the Kretschmann scalar (full contraction of the Riemann tensor) and is defined as 
\begin{align}
{\cal{K}}(r)&\equiv\sqrt{{\cal{R}}^{\mu\nu\rho\sigma}{\cal{R}}_{\mu\nu\rho\sigma}}
\nonumber\\
&= \bigg[(8\pi)^2\left \{3{\rho}^2(r)+3P^2(r)+2P(r){\rho}(r)\right\}
\nonumber\\
&
-\frac{128{\rho}(r)m(r)}{r^3} +\frac{48m^2(r)}{r^6}\bigg]^{1/2},
\label{KS}
\end{align}

At the surface, $m\rightarrow M$ as $r \rightarrow R$. Except for the last term, all other terms vanish outside the star because they depend on ${\rho}(r)$ and $P(r)$, which are zero outside the star. But, there is a non-vanishing component of the Riemann tensor that does not vanish; $\tensor{{\cal R}}{^1_{010}}=-\frac{2M}{R^3}=- \xi$, even in the outside of the star \cite{Kazim_2014, Xiao_2015}. Therefore, the Riemann tensor is a more relevant quantity to measure the curvature of the stars. Kretschmann scalar is the square root of the full contraction of the Riemann tensor. The vacuum value for $\cal{K}$ is
\begin{equation}
 \mathcal{K} =  \frac{4\sqrt{3} M}{R^3},
\end{equation}
as one can easily see from Eq. (\ref{KS}). Therefore, one can take $\cal K$ reasonable measures for the curvature within the star. The SC is defined as the ratio of curvature at the surface of the NS ${\cal{K}}(R)$ to the curvature of the Sun ${\cal{K}}_\odot$, SC $={\cal{K}}(R)/{\cal{K}}_\odot$. At the surface of the
Sun, the curvature is equal to $\cal{K}_\odot$ = 3.0 × 10$^{-17}$ km$^{-2}$ , which
can be considered as a small quantity. We show in Fig.~\ref{SC} the quantity $SC$ in units of $\times 10^{14}$.

We first notice that, despite the model utilized, the selected model, or even the anisotropy degree, all values of the $SC$ are in the same order of magnitude: $10^{14}$ times the curvature of the sun.
In relation to the isotropic case, we see that the $SC$ for the canonical star is ($\times 10^{14})$ 3.93 for the unpaired matter, 3.16 for the 2SC, and 2.94 for the CFL. When anisotropies are taken into account, as a positive value of $\lambda_{\rm BL}$ increases the radius of a fixed mass, we have a decrease of $SC$. For $\lambda_{\rm BL} = +2.0$, we have a decrease of around $8\%$ for the MIT, $7\%$ for the 2SC, and $4\%$ for the CFL. On the other hand, for $\lambda_{\rm BL} = -2.0$, we have an increase of around $10\%$ for the MIT and for the 2SC and $5\%$ for the CFL. For more massive strange stars, we see that anisotropy has a stronger influence.

Although today, there is no data for the $SC$ of neutron stars, as pointed out in Ref.~\cite{Kazim_2014}, this quantity can potentially act as a strong constraint in the future. For instance, in Ref.~\cite{Das_2021}, it was shown that the presence of dark matter in neutron star interiors can increase the curvature by one order of magnitude.
0.91-1.53
\begin{sidewaystable*}
 \scriptsize
\setlength\tabcolsep{7pt}
\begin{tabular}{|cc|cccccc|cccccccc|}
\hline
Model  & $\lambda_{\rm BL}$ & $M_{max}/M_\odot$   & $R_{1.4}$ (km) &$\Lambda_{1.4}$ & $I_{1.4}$  & $I_{max}$ & SC$_{1.4}$ & HESS & Capano & NICER$_{1.4}$ & J0740+6620 & J0952-0607 & GW170817 & GW190814 (M/$\Lambda$) & J0737-3039(A)\\
\hline
MIT &  -2.0 & 1.75 & 10.25 & 302 & 1.35 & 1.64 & 4.33 & \xmark & \xmark & \xmark & \xmark & \xmark & \checkmark & \xmark/\xmark & \checkmark   \\
MIT &  -1.5 & 1.83 & 10.35 & 319 & 1.38 & 1.77 & 4.22 & \xmark & \xmark & \xmark & \xmark & \xmark &  \checkmark & \xmark/\xmark & \checkmark  \\
MIT &  -1.0 & 1.91 & 10.44 & 340 & 1.41 & 1.93 & 4.11 & \xmark & \checkmark & \xmark & \xmark & \xmark & \checkmark &\xmark/\xmark & \checkmark  \\
MIT &  -0.5 & 2.00 & 10.52 & 361 & 1.44 & 2.14 & 4.02 & \xmark & \checkmark & \xmark & \xmark & \xmark & \checkmark & \xmark/\xmark & \checkmark \\
MIT &  0.0 & 2.09 & 10.60 & 383 & 1.47 & 2.36 & 3.93 & \xmark & \checkmark &  \xmark & \xmark & \xmark & \checkmark & \xmark/\xmark & \checkmark \\
MIT &  +0.5 & 2.19 & 10.67 & 406 & 1.49 & 2.64 & 3.85 & \checkmark & \checkmark & \xmark & \xmark & \xmark & \checkmark & \xmark/\xmark & \checkmark \\
MIT &  +1.0 & 2.31 & 10.74 & 431 & 1.52 & 2.99 & 3.78 & \checkmark & \checkmark & \xmark & \checkmark & \checkmark & \checkmark  & \xmark/\xmark & \checkmark   \\
MIT &  +1.5 & 2.44 & 10.81 & 456 & 1.55 & 3.39 & 3.71 & \checkmark & \checkmark & \xmark & \checkmark & \checkmark & \checkmark & \xmark/\xmark  & \checkmark  \\
MIT &  +2.0 & 2.58 & 10.87 & 482 & 1.57 & 3.89 & 3.64 & \checkmark & \checkmark & \xmark & \checkmark & \checkmark & \checkmark & \checkmark/\checkmark & \checkmark    \\
\hline
2SC &  -2.0 & 1.80 & 11.05 & 495 & 1.53 & 1.88 & 3.47 & \checkmark  & \checkmark  & \xmark & \xmark  & \xmark & \checkmark & \xmark/\checkmark & \checkmark \\
2SC &  -1.5 & 1.88 & 11.14 & 523 & 1.57 & 2.07 & 3.38  & \checkmark & \checkmark & \xmark & \xmark & \xmark & \checkmark & \xmark/\checkmark & \checkmark  \\
2SC &  -1.0 & 1.97 & 11.23 & 555 &  1.60 & 2.28 & 3.30 & \checkmark & \checkmark & \xmark & \xmark & \xmark & \checkmark & \xmark/\checkmark & \checkmark\\
2SC &  -0.5 & 2.06 & 11.31 & 588 & 1.63 & 2.53 & 3.24  & \checkmark & \checkmark & \xmark  & \checkmark & \xmark & \xmark & \xmark/\checkmark & \checkmark\\
2SC &  0.0 & 2.16 & 11.39 & 622 & 1.66 & 2.82 & 3.16   & \checkmark & \checkmark & \xmark & \checkmark & \xmark & \xmark & \xmark/\checkmark & \checkmark\\
2SC &  +0.5 & 2.27 & 11.46 & 658 & 1.69 & 3.17 & 3.10  & \checkmark & \checkmark & \xmark & \checkmark & \checkmark & \xmark & \xmark/\checkmark & \checkmark \\
2SC &  +1.0 & 2.40 & 11.53 & 696 & 1.72 & 3.57 & 3.05  & \checkmark & \checkmark & \xmark & \checkmark & \checkmark & \xmark & \xmark/\checkmark &  \checkmark \\
2SC &  +1.5 & 2.54 & 11.60 & 733 & 1.75 & 4.07 & 2.99  & \checkmark & \checkmark & \xmark & \checkmark & \checkmark & \xmark & \checkmark/\checkmark  & \xmark\\
2SC &  +2.0 & 2.70 & 11.66 & 775 & 1.77 & 4.71 & 2.94  & \checkmark & \checkmark & \xmark & \checkmark & \checkmark & \xmark & \checkmark/\checkmark & \xmark\\
\hline
CFL &  -2.0 & 2.32 & 11.49 & 699 & 1.70 & 3.32 & 3.08 & \checkmark & \checkmark & \xmark & \checkmark & \checkmark & \xmark & \xmark/\checkmark & \checkmark\\  
CFL &  -1.5 & 2.42 & 11.54 & 718 & 1.72 & 3.64 & 3.04  & \checkmark & \checkmark & \xmark & \checkmark & \checkmark & \xmark & \xmark/\checkmark & \checkmark\\  
CFL &  -1.0 & 2.52 & 11.59 & 739&  1.74 & 3.98 & 3.00   & \checkmark & \checkmark & \xmark & \checkmark & \checkmark & \xmark & \checkmark/\checkmark & \checkmark\\  
CFL &  -0.5 & 2.63 & 11.63 & 762 & 1.76 & 4.41 & 2.97  & \checkmark & \checkmark & \xmark & \checkmark & \checkmark & \xmark & \checkmark/\checkmark & \xmark\\  
CFL &  0.0 & 2.75 & 11.67 & 787 & 1.78 & 4.92 & 2.94  & \checkmark & \checkmark & \xmark & \checkmark & \checkmark & \xmark & \checkmark/\checkmark & \xmark\\  
CFL &  +0.5 & 2.89 & 11.71 & 812 & 1.79 & 5.48 & 2.91  & \checkmark & \checkmark & \xmark & \checkmark & \checkmark & \xmark & \checkmark/\checkmark & \xmark\\  
CFL &  +1.0 & 3.03 & 11.75 & 838 & 1.81 & 6.14 & 2.89 & \checkmark & \checkmark & \xmark & \checkmark & \checkmark & \xmark & \checkmark/\checkmark & \xmark\\    
CFL &  +1.5 & 3.19 & 11.78 & 856 & 1.83 & 6.95 & 2.86  & \checkmark & \checkmark & \xmark & \xmark & \checkmark & \xmark & \checkmark/\checkmark & \xmark\\  
CFL &  +2.0 & 3.37 & 11.82 & 899 & 1.84 & 7.95 & 2.83    & \checkmark & \checkmark & \xmark & \xmark & \checkmark & \xmark & \checkmark/\xmark & \xmark\\  
\hline
\end{tabular}
\caption{Some strange star properties for unpaired (MIT), 2SC, and CFL quark matter and some constraints discussed in the text. The inertia moment is given in units of  $\times~10^{45}$~g.cm$^{-2}$.  The scalar surface,  SC $={\cal{K}}(R)/{\cal{K}}_\odot$, is given in unit of $10^{14}$.}
\label{T1}
\end{sidewaystable*}

The main results of this study, as well as some constraints discussed, are summarized in Tab.~\ref{T1}.
Before we finished, we quickly analyzed some results presented in this table. As pointed out before, our main goal is to fulfill three constraints that are not easily fulfilled simultaneously in ordinary hadronic models: the small radius of the canonical star ($R_{1.4}~< 11.9$ km) presented in Ref.~\cite{Capano_2020}; the existence of very light compact objects, HESS J1731-347~\cite{Doroshenko_2022}; and the speculative mass of the black widow pulsar PSR J0952-0607~\cite{Romani_2022}.

The models that satisfy simultaneously these three constraints are the MIT for $\lambda_{\rm BL}~\geq +1.0$, 2SC for $\lambda_{\rm BL}~\geq +0.5$, and the CFL for all degrees of anisotropy. If altogether with those three bounds, we impose that the model also needs to fulfill the mass and radius range of the PSR J0740+6620~\cite{Miller_2021} and the inertia moment of the J0737-3039(A)~\cite{Landry_2018}, therefore we must rule out the CFL for $\lambda_{\rm BL}~\geq -0.5$, while for the 2SC, only the values $\lambda_{\rm BL}$ = +0.5 and +1.0 are now valid. 
On the other hand, if altogether with the three main bounds, we impose that the model must satisfy the constraint related to the dimensionless tidal parameter in the GW170817 event. Therefore, we must rule out all models based on super-conducting quark matter. We finish by pointing out that unpaired quark matter with $\lambda_{\rm BL} = +2.0$ virtually fulfills every constraint presented. The only exception is the radius constraint for the canonical star. However, we must emphasize that the window comprehending the NICER ($R_{1.4}~>11.8$ km)~\cite{Miller_2021} and the Capano ($R_{1.4}~< 11.9$ km)~\cite{Capano_2020} is very narrow.

\section{Conclusion}
In this work, we study the effect of the anisotropy in strange stars within three different quark models: unpaired quark matter, 2SC, and CFL. For the unpaired quark model, we use the vector MIT bag model as introduced in Ref.~\cite{Lopes_2021, Lopes_2021b}. For both super-conducting quark matter (2SC and CFL) we use the CSS model~\cite{Zdunik2013, Alford2015, Burgio2016, Alford2013, Han2019, Cierniak_2020, Sun2023, Lopes2023JCAP}. The parameters for the 2SC and CFL are taken from Ref.~\cite{Zdunik2013}, and as an additional constraint, we impose that the unpaired quark matter must be softer than the color super-conducting one, as indicated in the literature~\cite{Alford2008, Lugones2004}. The effects of the anisotropy are studied in the BL model~\cite{Bowers_1974}, which has the advantage of preserving the spherical symmetry of the star~\cite{DelgadoeDelgado2018}.

We searched for models that are able to satisfy three strong constraints simultaneously: the small radius of the canonical star ($R_{1.4}~< 11.9$ km) presented in Ref.~\cite{Capano_2020}; the existence of very light compact objects, HESS J1731-347~\cite{Doroshenko_2022}; and the speculative mass of the black widow pulsar PSR J0952-0607~\cite{Romani_2022}. Additional constraints, as well as additional features such as the dimensionless tidal parameter, $\Lambda$, the MOI, and SC are also presented. The main results are summarized below:
\begin{itemize}

    \item In the mass-radius diagram, we observed that the presence of the anisotropy increases the maximum mass, as well as the radius of a fixed mass star for positive values of $\lambda_{\rm BL}$, and vice-versa for negative ones.

    \item The effect of anisotropies in the maximum mass is similar for all three quark matter models; however, the effect in the canonical star is stronger in the softer models.

    \item The models that are able to fulfill the three main constraints simultaneously are the MIT for       $\lambda_{\rm BL}~\geq +1.0$, 2SC for $\lambda_{\rm BL}~\geq +0.5$, and the CFL for all degrees of anisotropy. 

     \item The anisotropy increase the tidal dimensionless parameter $\Lambda$ for positive values of $\lambda_{\rm BL}$ and reduce it for negative ones. Despite this, only unpaired quark matter can fulfill the constraints related to the GW170817 event.

     \item The MIT  presents the smaller values of MOI, and the CFL model has the large MOI. The anisotropies change the MOI of the canonical star by around $10\%$. However, due to the proportionality of $I$ with $MR^2$, the anisotropy strongly affects the maximum mass of strange stars. The MOI increases more than 60$\%$ for positive $\lambda_{\rm BL}$ values and  decreases around 30$\%$ for negative values. Only for the MIT the constraints related to the J0737-3039(A), are satisfied for all values of $\lambda_{\rm BL}$.

     \item The anisotropy has a weak influence of the SC in low-mass strange stars and a moderate one in most massive ones. For all values of mass and all degrees of anisotropy, the SC is always in the same magnitude order ($10^{14}$). 

     \item The unpaired matter with $\lambda_{\rm BL} = +2.0$ virtually fulfill all constraints presented, with the exception of the revised NICER constraint.

     \item Therefore, it has been observed that pressure anisotropy has significant effects on the various strange star properties. There are still unsolved issues based on the origin of anisotropy and its effects on the various properties of compact stars, which will be discussed in future work.
\end{itemize}
\textbf{Acknowledgements:} L.L.L. was partially supported by CNPq Universal Grant No. 409029/2021-1.
\bibliography{ASQS}
\bibliographystyle{spphys}
\end{document}